\documentclass{PoS}
\usepackage{epsfig}
\usepackage{amssymb}
\usepackage{amsfonts}
\usepackage{subfigure}
\title{Ergodic SO(3), monopole condensation and vortex free energy}

\ShortTitle{Ergodic SO(3), monopole condensation and vortex free energy}

\author{\speaker{Giuseppe Burgio}\\
        Humboldt-Universit\"at zu Berlin\\
        E-mail: \email{burgio@physik.hu-berlin.de}}

\author{Marcel Fuhrmann\\
        Humboldt-Universit\"at zu Berlin\\
        E-mail: \email{fuhrmann@physik.hu-berlin.de}}

\author{Werner Kerler\\
        Humboldt-Universitaet zu Berlin\\
        E-mail: \email{kerler@physik.hu-berlin.de}}

\author{Michael M\"uller-Preussker\\
        Humboldt-Universitaet zu Berlin\\
        E-mail: \email{mmp@physik.hu-berlin.de}}

\abstract{We study the continuum limit of adjoint SU(2) LGT by means 
of a suppression term for $\mathbb{Z}_2$ monopoles. High
barriers for tunnelling among different twist sectors are overcome through
pa\-ral\-lel tempering. Monopole condensation is used
to study the deconfinement transition and the properties of the confined
phase. Ergodicity in summing over all twist sectors allows an unbiased measure 
of the 't Hooft vortices free energy. Its behaviour in the SO(3) 
confined phase hints at differences from what conjectured for semi-integer
discretizations.}

\FullConference{XXIIIrd International Symposium on Lattice Field Theory\\
		 25-30 July 2005\\
		 Trinity College, Dublin, Ireland}

\begin{document}

\section{Introduction}

The spontaneous breaking of center simmetry 
\cite{Polyakov:1978vu}
% \cite{Susskind:1979up}
observed in simulations of pure fundamental lattice Yang-Mills theories has 
offered unique insight in their still elusive dynamics
\cite{McLerran:1981pk}. 
% \cite{Kuti:1981gh}. 
Whether and in what sense this holds for the perturbatively 
equivalent center-blind adjoint discretization,
as universality predicts \cite{Svetitsky:1982gs}, must still be 
appropriately answered \cite{Smilga:1993vb}. The difficulties inherent 
to such non-perturbative regularization have been well known for a long 
time \cite{Greensite:1981hw}
and are best illustrated by the phase diagram of the 
mixed action,
% \cite{Bhanot:1981eb}
which for SU(2) reads
\begin{equation}
S = \beta_{A}\sum_{P}\Bigg(1-\frac{1}{3}Tr_{A}U_{P}\Bigg)+\beta_{F}\sum_{P}
\Bigg(1-\frac{1}{2}Tr_{F}U_{P}\Bigg)\,;\;\;
\frac{1}{g^2} =  \frac{1}{4}\beta_F+\frac{2}{3}\beta_A\,.
\label{mixed}
\end{equation}
The theory exhibits bulk transitions related to the
condensation of $\mathbb{Z}_2$ monopoles $\sigma_{c}$ $\in$ SO(3)
and vortices $\sigma_{l}$ $\in$ SU(2) which hinder the study
of its finite temperature properties \cite{Halliday:1981te}.
% \cite{Halliday:1981tm}.
First concrete attempts to study itxsy at finite temperature by 
implementing suppressing chemical potentials $\lambda\sum_{c}(1-\sigma_{c})$, 
$\gamma\sum_{l}(1-\sigma_{l})$, as suggested in \cite{Halliday:1981te},
% \cite{Halliday:1981tm}, 
were only made relatively recently
\cite{Datta:1997nv}.
For the center blind case $\beta_F=0, \gamma=0$, where the bulk transition 
separates a strong coupling
phase I, continuosly connected with SU(2), from a weak coupling phase II, 
there is no simmetry breaking mechanism and no order parameter.
In these works it was however first observed
how in such pure adjoint case at high temperature the theory possesses, 
besides the 
``regular'' deconfined phase where the adjoint Polyakov loop $L_A \to 1$,
a new phase where $L_A \to -1/3$. In \cite{deForcrand:2002vs} a
dynamical observable measuring the twist expectation value $z$ was introduced 
after noting 
that the $\delta(\sigma_c-1)$ constraint effectively implemented by 
$\mathbb{Z}_2$ monopole suppression 
allows the SO(3) partition function to be rewritten
as the sum of SU(2)
partition functions with all possible twisted boundary conditions 
$Z|_{z={\rm i}}$
\cite{Mack:1979gb}.
%\cite{Tomboulis:1980vt}
The $L_A \to -1/3$ phase was linked to a 
non trivial
twist expectation value, i.e. to the creation of a vortex in the vacuum. 
The SO(3) theory was proposed as the ideal test case
to check the 't~Hooft vortex confinement criterion 
\cite{'tHooft:1977hy}.
% \cite{'tHooft:1979uj}. 
Unfortunately twist sectors 
freeze at the bulk and attempts with a multicanonical algorythm
at $\lambda=0$ have been limited to volumes 
not higher then $8^3 \times 4$ \cite{deForcrand:2002vs}.
Other attempts to study 
the existence
of a finite temperature transition in phase II with $\lambda \neq 0$
through thermodynamic
observables were limited to very small temporal extent $N_\tau=2$ 
\cite{Datta:1999di},
% \cite{Datta:1999np}.
confirming that the coninuum limit is a real challenge for the adjoint theory.
A step forward was made in \cite{Barresi:2004qa,Barresi:2006gq},
where by means of the Pisa disorder parameter
for monopole condensation lines of second order transition properly 
scaling with $N_\tau$ and ending on the bulk 
where actually
found at each {\em fixed} twist, with critical exponents consistent
with Ising 3-d. 
Whether such
is the case also for the theory summed over all twist sectors and how the
vortex free energy behaves in the ergodic simulations is the subject 
of the present preliminary report, based on the poster presented
at this conference. A complete analysis of the model with updated results,
including a detailed
description of the algorythm and error analysis can be found in \cite{Burgio:2006dc,Burgio:2006xj}. 

\section{Action and Observables}
As anticipated,
we will concentrate on the pure adjoint Wilson action 
(Eq.~(\ref{mixed}) with $\beta_F =0$),
modified by the suppression term $\lambda\sum_{c}(1-\sigma_{c})$, where
%\begin{eqnarray} 
%S=\frac{4}{3}\beta_{A} \sum_{P} 
%  \left(1-\frac{1}{4}\mathrm{Tr}_{F}^{2}U_{P}\right)
%  +\lambda \sum_{c}(1-\sigma_{c})\,.
%\label{ouraction}
%\end{eqnarray}
$\sigma_{c}=\prod_{P\in\partial c}\mathrm{sign}(\mathrm{Tr}_{F}U_{P})$
around all elementary 3-cubes $c$ defines the 
$\mathbb{Z}_2$ magnetic charge. Its density 
$M = 1-\langle\frac{1}{N_c}\sum_{c}\sigma_{c}\rangle$
tends to one in the strong coupling region (phase I)
and to zero in the weak coupling limit (phase II), $N_c$ denoting the total 
number of elementary 3-cubes.  
Such action is center-blind
in the entire $\beta_A-\lambda$ plane \cite{Barresi:2003jq}. 
We will employ parallel tempering to obtain
ergodicity among different twist sectors when evaluating the expectation
values of physical observables, e.g. the Pisa disorder parameter and the
't~Hooft vortex free energy. 

\subsection{Vortex Free Energy}
Temporal twists, corresponding to maximal 't~Hooft loop and defining our 
center vortices, are topological excitations which offer a natural link 
between center simmetry breaking and degrees of freedom independent
of the discretization used \cite{'tHooft:1977hy}. 
It has been known for a long time in the literature that SO(3) with 
$\mathbb{Z}_2$ monopole chemical potential in phase II is equivalent to 
SU(2) including all possible twisted b.c. 
\cite{Mack:1979gb}
% \cite{Tomboulis:1980vt}
\begin{equation}
\sum_{\mathrm{b.c.}}Z_{\sf SU(2)}
=\int (DU) 
e^{-S_{\sf SO(3)}}
\prod_c \delta(\sigma_c-1) \simeq Z_{\sf SO(3)}\mid_{\lambda \to \infty}.
\end{equation}
Above the bulk SO(3) trades thus boundary conditions with 
twist sectors. Twists, i.e. 't~Hooft loops, 
thus become {\it observables} rather than boundary 
constraints like in SU(2):
\begin{equation}
z_i = \frac{1}{N_s^2}\sum_{j,k\neq i} \prod_{x \in (i,t) {\sf plane}} 
{\sf sign}({\sf Tr}_{f}U_{i,t}(x))\,.
\end{equation}
Since creating such 't~Hooft 
loop amounts to a change in the
signs of some plaquettes, the free energy change 
$\Delta F=\Delta U-T \Delta S$ 
will only receive an entropy contribution for adjoint discretizations, 
the action remining unmodified in the process. Definying thus the 't~Hooft 
vortex free energy as the 
ratio of the partition function in the non-trivial twist sector to that in the 
trivial one $F = -T \log{{Z|_{z=1}}/{Z|_{z=0}}}$, their relative weight can be measured through an egodic 
simulation. To be an order parameter in the 
thermodynamic limit 
($V=N_s^3\to\infty$),
$F$ should vanish exponentially in the confined phase while diverging with 
an area law 
$F\sim \tilde{\sigma}N_s^{2}$ above the deconfinement transition, where
$\tilde{\sigma}$ is the dual string tension. 
Working at $\lambda=0$ proved
however to be a hurdle, since the ``freezing'' of twist sectors above the bulk 
transition 
yields high potential barriers hard to overcome even with a 
multi-canonical
algorythm \cite{deForcrand:2002vs}.
%Twist sectors were shown to be well defined in all phase II indepentently 
%of $L_A$
%\cite{Barresi:2001dt,Barresi:2002un}, the latter actually still approximately 
%\cite{Smilga:1993vb,Michael:1985ne}
%satisfying an Haar-measure distribution
%above the 2$^{\rm nd}$ order branch of the bulk transition and
%departing from it at different $\beta_A^c$ for
%different $N_\tau$
%as $\beta_A$ increases \cite{Barresi:2003jq}. 
%This hints at a transition line from a confined 
%to a deconfined phase in each {\it fixed} twist sector 
%\cite{Barresi:2003jq,Barresi:2004gk,Barresi:2003yb} ending on the
%bulk transition.
From the results in 
\cite{Barresi:2003jq,Barresi:2001dt,Barresi:2002un,Barresi:2004gk,Barresi:2003yb}
it is sound to conjecture that the whole physically relevant
SO(3) dynamics lies 
in phase II, the finite temperature transition eventually decoupling from
the bulk even at $\lambda=0$. Unfortunately, from
estimates in \cite{deForcrand:2002vs}, this should not happen
for volumes smaller than $\sim 500 \times 1000^3.$ A non-vanishing
$\lambda$ seems therefore the only feasible way to gain access
to the properties of the continuum limit of SO(3). The observation 
made in \cite{Datta:1997nv} that the bulk transition 
weakens to 2$^{\rm nd}$
order with increasing $\lambda$ will prove crucial in this respect. 

\subsection{Pisa disorder parameter}

The Pisa disorder operator, measuring monopole condensation, is an 
order parameter for the dual superconductor mechanism of confinement 
\cite{DiGiacomo:1997sm}:
% \cite{DiGiacomo:1999fa}:
defined through $S_M$, the action modified through a bosonic field 
$\Phi$ introduced at fixed time and invariant 
under a residual gauge simmetry U(1). It is independent of the particular 
choice for $\Phi$ and is well defined, independently of center simmetry, 
also for full QCD. Its derivative, $\rho$, is easier to compute in actual numerical simulations.
\begin{equation}
\langle \mu\rangle =
\frac{\int (DU) e^{-(S_M-S)}e^{-S}}{\int (DU) e^{-S}}=\exp\left(\int_0^\beta \rho(\beta')d\beta'\right)
\end{equation}
For $T<T_c$ $\langle \mu \rangle \neq 0$ 
signals spontaneous breaking of U(1), corresponding to $\rho \sim 0$, bounded 
from below for $N_s \to \infty$. At the phase transition $\rho$ should show a 
sharp negative peak at $\beta^c$ diverging for $N_s \to \infty$, while for 
$T>T_c$ $\langle \mu \rangle = 0 $ corresponds to the trivial vacuum ($\rho 
\propto - N_s$ for $N_s \to \infty$). More details regarding its 
implementation for SO(3) can be found in \cite{Barresi:2004qa,Barresi:2006gq}.

\section{Fixed twist dynamics}
\label{fix}
It was shown in \cite{Barresi:2006gq} that for SO(3) in phase II at fixed twist 
at low $\beta_A$ $\rho$ $\sim 0$ $\forall z$, i.e. there always exhists 
a confined phase; $\rho$ peaks at some ($z$ dependent!) $\beta_A^c$;
at high $\beta_A$ $\rho$ diverges for $z=0$, while it vanishes for $z\neq 0$.
Moreover each $z$ sector can be mapped in a different positive plaquette 
model. This makes fundamental observables measurable, while the adjoint sector 
remains untouched. It was thus proven that different twists possess 
slightly different underlying 
fundamental dynamics in the confined phase, but very different ones 
above deconfinement, the twist behaving effectively 
like a background field. Each twist sector measures slightly different 
string tensions in the confined phase (through Creutz ratios and
Polyakov loop correlators) and deconfinement temperatures. 
From the above fixed twist measurements it is thus straightforward to see 
that the ergodic expectation value of $\mu$
\begin{equation}
\langle\mu\rangle = \frac{\sum_{ i} \mu|_{z={ i}} Z_{ SO(3)}|_{z={ i}}}{\sum_{ i} Z_{ SO(3)}|_{z={ i}}}
\end{equation}
at low $\beta_A$ will be $\langle\mu\rangle \simeq 0$, while 
at high $\beta_A$ $\langle\mu\rangle \simeq \langle\mu\rangle|_{z=0} (1-e^{-{F}/{T}})$. 
In whole phase II ergodic theory confines at low $\beta_A$, deconfines at high 
$\beta_A$. To establish whether $\mu$ actually is an order parameter its
behaviour at the physical transition must be studied through an ergodic 
algorythm.

\section{Parallel Tempering}

\begin{figure}[ht]
\begin{center}
\includegraphics[width=7cm]{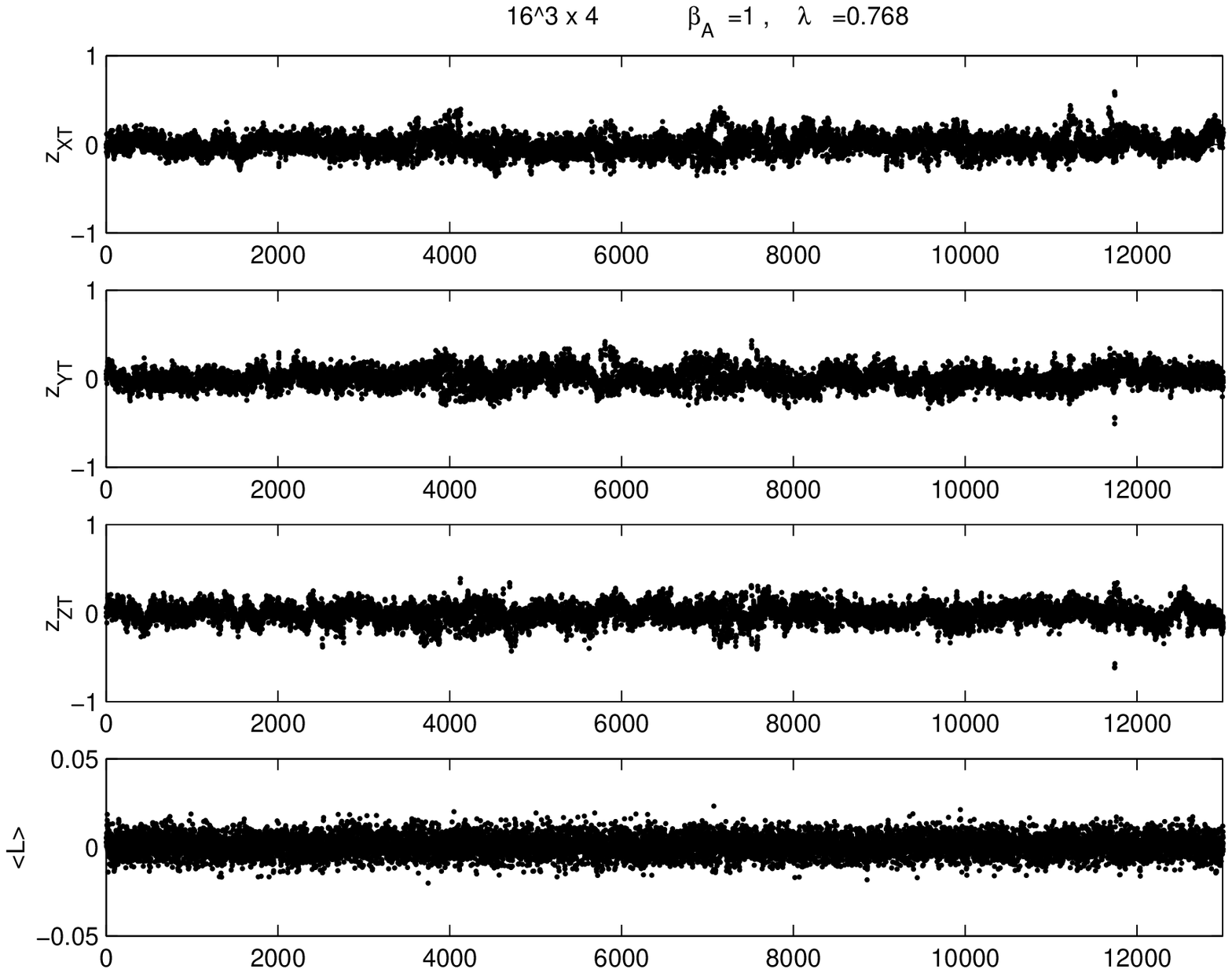}
\includegraphics[width=7cm]{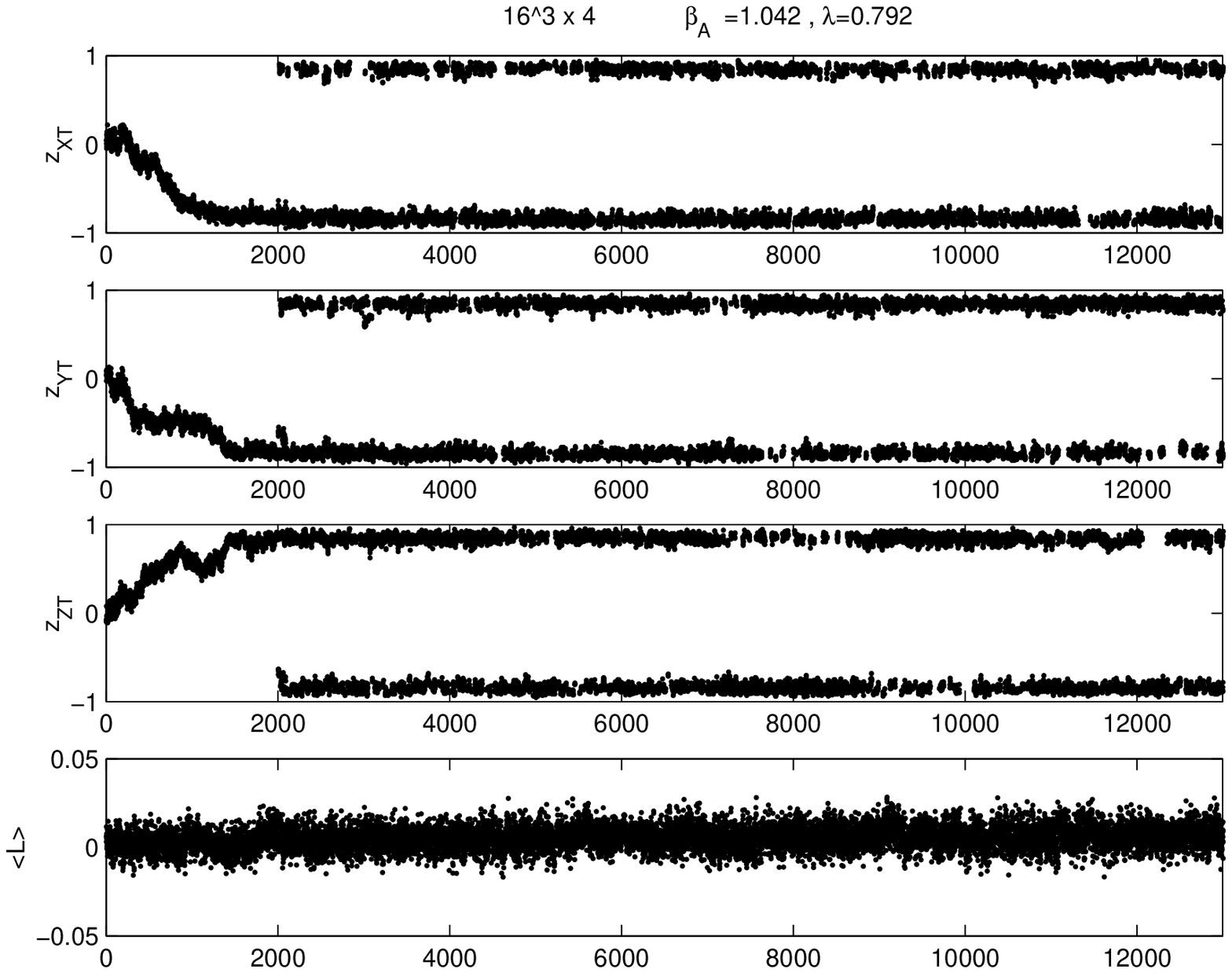}
\end{center}
\caption{MC history of twists and $L_A$ on top (left) and above bulk (right), 
$V=16^3\times 4$.}
\label{par}
\end{figure}

The idea of Parallel tempering (see \cite{Marinari:1996dh} for a review)
is to simulate at the same time many ensambles above and below the
bulk transition. Each ensemble is caracterized by a link-configuration 
$C_i$ and a set of couplings $\beta_i$. 
To reach ergodicity one lets them evolve separately swapping
every few MC steps the $i^{ th}$ and $(i-1)^{ th}$ ensemble through a
generalized Metropolis step with weight:
\begin{equation}
W_i  = e^{-(S(\beta_i,C_{i-1})+S(\beta_{i-1},C_{i}))+(S(\beta_i,C_{i})+S(\beta_{i-1},C_{i-1}))}.
\end{equation}
As Fig.~\ref{par} shows it works very well, 
although it is crucial to fine tune the parameters to 
optimize acceptance rate and keep
auto- and cross-correlations under control \cite{Burgio:2006dc,Burgio:2006xj}. 
Since phase I has a high tunnelling probability, the softening of the bulk 
transition to 2$^{\rm nd}$ order is crucial to "transport" tunneling into 
phase II. Working on the 1$^{\rm st}$ 
order bulk branch would lead to high barriers and thus no ergodicity 
at high volume.

\section{Results}

\begin{figure}[ht]
\begin{center}
\includegraphics[width=7cm]{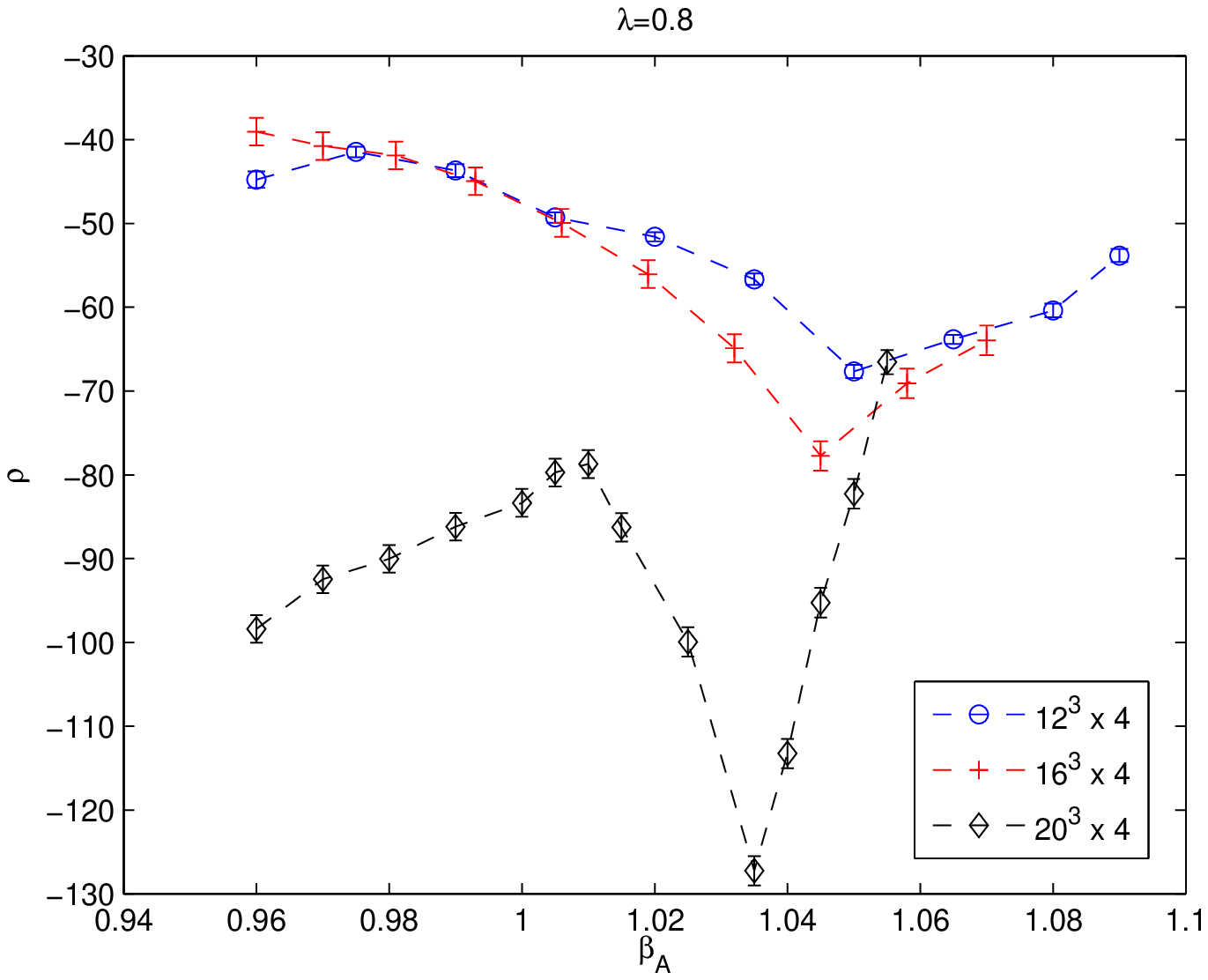} 
\includegraphics[width=7cm]{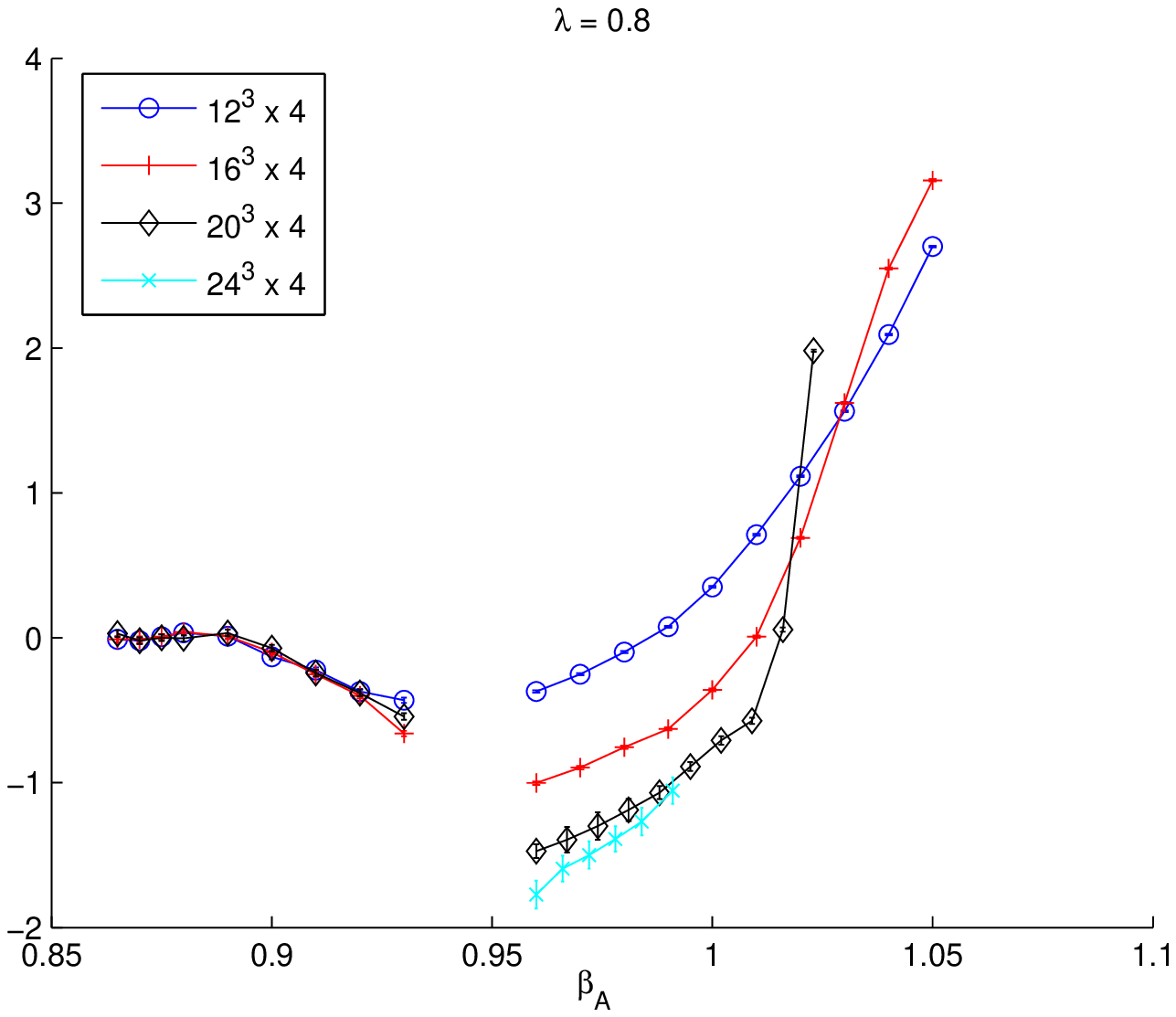}
\end{center}
\caption{Pisa disorder operator (right) and vortex free energy (left).}
\label{res}
\end{figure}

As explained in Sect.~(\ref{fix}), we need to establish whether 
$\rho$ gives a clear signal for a phase transition through ergodic runs. 
This is shown in Fig.~(\ref{res}-right). Due to the algorythm, 
we must choose a path that starts close to the bulk and 
goes a\-bove the de\-con\-fi\-ne\-ment trans\-it\-ion, with both 
not too far away from each other.
Such path gives competing effects, since 
$\rho$ also diverges at the bulk \cite{Barresi:2004qa, Barresi:2006gq}.
As for $F$, in phase I and at high $\beta_A$ it behaves as expected.
The surprise comes in the confined phase of phase II, where the
free en\-er\-gy doesn't va\-nish (Fig.~\ref{res}-left)
We checked that away from both transitions $F$ 
reaches constant, $\beta_A$ dependent  
negative values throughout the confined phase of phase II.

\section{Conclusion and Outlook}

We study the finite temperature transition in the continuum limit of SO(3) 
lattice gauge theory.
Sum over twists for ergodicity is reached through Parallel tempering.
The Pisa disorder pa\-ra\-me\-ter is used to cha\-ract\-erize the 
pro\-per\-ties
of the different phases, so that Mo\-no\-pole (de)co\-nden\-sa\-tion 
could be used to obtain the critical exponents.
Although well defined for SO(3), the 't~Hooft vortex free energy behaves 
unexpectedly, being negative in the confined phase of SO(3).  
The meaning of such vortex enhancement could be relevant for the dynamics of
continuum Yang-Mills theories. 
However, contrary to $\mu$, $F$ seems not to behave as an order parameter 
in the theory discretized with\-out
fundamental components.
Keeping in mind the full QCD case, there actually 
seems to be no compelling reason 
why in a center-blind regularization of Yang-Mills theory 
an order parameter for 
center simmetry breaking should vanish. Universal, representation 
independent observables should however be preserved, so further confirmations 
of the correct physical behaviour of the adjoint dynamics through 
e.g. the glueball spectrum 
might therefore be interesting. Also, the extension to SU(3) should
shed further light on the problem. For further discussions and a more 
detailed analysis refer to \cite{Burgio:2006dc,Burgio:2006xj}.

\end{document}